\documentclass[final,english]{bullsrsl}[2022/06/15]

\usepackage[latin1]{inputenc}
\usepackage[T1]{fontenc}

\usepackage{natbib} 
\usepackage{graphicx}

\begin{document}
\title{The 4m International Liquid Mirror Telescope: a brief history and some preliminary scientific results}

\author[affil={1}, corresponding]{Jean}{Surdej}
\author[affil={2,3}]{Bhavya}{Ailawadhi}
\author[affil={4,5}]{Talat}{Akhunov}
\author[affil={6}]{Ermanno}{Borra}
\author[affil={2,7}]{Monalisa}{Dubey}
\author[affil={2,7}]{Naveen}{Dukiya}
\author[affil={8}]{Jiuyang}{Fu}
\author[affil={8}]{Baldeep}{Grewal}
\author[affil={8}]{Paul}{Hickson}
\author[affil={2}]{Brajesh}{Kumar}
\author[affil={2}]{Kuntal}{Misra}
\author[affil={2,3}]{Vibhore}{Negi}
\author[affil={1}]{Anna}{Pospieszalska-Surdej}
\author[affil={2,9}]{Kumar}{Pranshu}
\author[affil={8}]{Ethen}{Sun}
\affiliation[1]{Institute of Astrophysics and Geophysics, University of Li\`{e}ge, All\'{e}e du 6 Ao$\hat{\rm u}$t 19c, 4000 Li\`{e}ge, Belgium}
\affiliation[2]{Aryabhatta Research Institute of observational sciencES (ARIES), Manora Peak, Nainital, 263001, India}
\affiliation[3]{Department of Physics, Deen Dayal Upadhyaya Gorakhpur University, Gorakhpur, 273009, India}
\affiliation[4]{National University of Uzbekistan, Department of Astronomy and Astrophysics, 100174 Tashkent, Uzbekistan}
\affiliation[5]{ Ulugh Beg Astronomical Institute of the Uzbek Academy of Sciences, Astronomicheskaya 33, 100052 Tashkent, Uzbekistan}
\affiliation[6]{Department of Physics, Universit\'{e} Laval, 2325, rue de l'Universit\'{e}, Qu\'{e}bec, G1V 0A6, Canada}
\affiliation[7]{Department of Applied Physics, Mahatma Jyotiba Phule Rohilkhand University, Bareilly, 243006, India}
\affiliation[8]{Department of Physics and Astronomy, University of British Columbia, 6224 Agricultural Road, Vancouver, BC V6T 1Z1, Canada}
\affiliation[9]{Department of Applied Optics and Photonics, University of Calcutta, Kolkata, 700106, India}


\correspondance{jsurdej@uliege.be}
\date{6th May 2023}
\maketitle


%

\begin{abstract}

The present article is based upon an invited talk delivered at the occasion of the inauguration of the  4m International Liquid Mirror Telescope (ILMT) which took place in Devasthal (ARIES, Uttarakhand, India) on 21st of March 2023. We present hereafter a short history of the liquid mirror telescopes and in particular of the 4m ILMT which is the first liquid mirror telescope entirely dedicated to astrophysical observations. We discuss a few preliminary scientific results and illustrate some direct CCD images taken during the first commissioning phase of the telescope. We invite the reader to refer to the series of ILMT poster papers published in these same proceedings of the BINA3 workshop for more details about the instrument, operation, first observations, performance and scientific results.

\end{abstract}

\keywords{ILMT, survey, telescope, inauguration, first light}


\section{A brief history}

The concept of a liquid mirror telescope can be traced back to the XVIIth century when Isaac Newton  demonstrated that the surface of a spinning liquid takes the shape of a paraboloid, the perfect surface of reference to focus into a single point a beam of parallel  light rays.

However, it was not until 1850 that an Italian astronomer, Ernesto Capocci - then director of Naples$^\prime$s observatory, revived the idea to use a rotating container filled with mercury as the primary mirror of an astronomical telescope (\citealt{paper9}). But at that time, the concept was not taken seriously, mainly because a mercury mirror cannot be tilted to track moving objects. Indeed,
because of the Earth$^\prime$s rotation, astronomical objects would produce long streaks instead of stellar images on the photographic emulsions previously used in astronomy.  Let us immediately point out that the advent of charge coupled devices (CCDs) in the 1970s changed the situation. CCDs enable a technique, called time-delayed integration (TDI), to compensate for the Earth's rotation (\citealt{paper3}). This is done by electronically shifting the electronic charges on the surface of the CCD at the same speed as the stellar images drift in the focal plane of the telescope, thereby yielding sharp images. TDI has revived scientific interest in liquid mirrors.  

But it was not until the end of the XIXth century that anyone tried to build a liquid mirror (LM). In 1875, Henry Skey (New Zealand) built in the laboratory a $\sim$ 35cm LM.  In 1909, Robert Wood constructed at the Johns Hopkins University (Baltimore, USA) a 51 cm prototype, and made first sky observations but annoying ripples were produced at the surface of the mercury because of vibration transmission and the difficulty to keep constant the angular rotation of the mirror. 

Starting in the early 1980s, scientists mostly in Canada have been engaged in significant development work in the laboratory and observatories. Researchers first demonstrated the feasibility of large liquid optics and developed the basic technology behind it. Optical shop tests of liquid mirrors with diameters as large as 2.5-m had shown diffraction-limited optical quality. 

For instance, in 1982 Ermanno Borra found  a solution to some of the  technical challenges encountered by Robert Wood. The former suggested dampening the vibrations that caused the ripples by using a pressurized air-bearing to sustain the dish.  He also suggested pouring a liquid resin on the dish surface first, letting it dry into the right shape, then pouring reflective liquid onto it as a coating, diminishing at the same time the amount of mercury needed. 

\begin{figure}
\centering
\includegraphics[width=12cm]{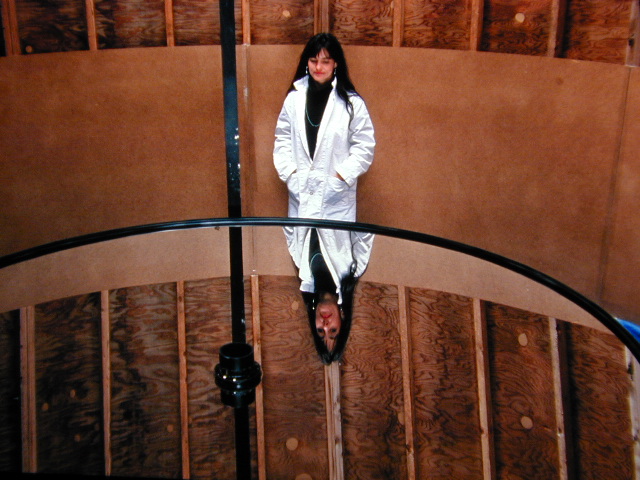}\label{Fig_1}
\begin{minipage}{12.0cm}
\caption{The UBC-Laval 2.7-m liquid mirror built in 1994 (\copyright \, Prof. Paul Hicskon).}
\end{minipage}
\end{figure}

\begin{figure}
\centering
\begin{minipage}{6.0cm}
\includegraphics[width=6.0cm]{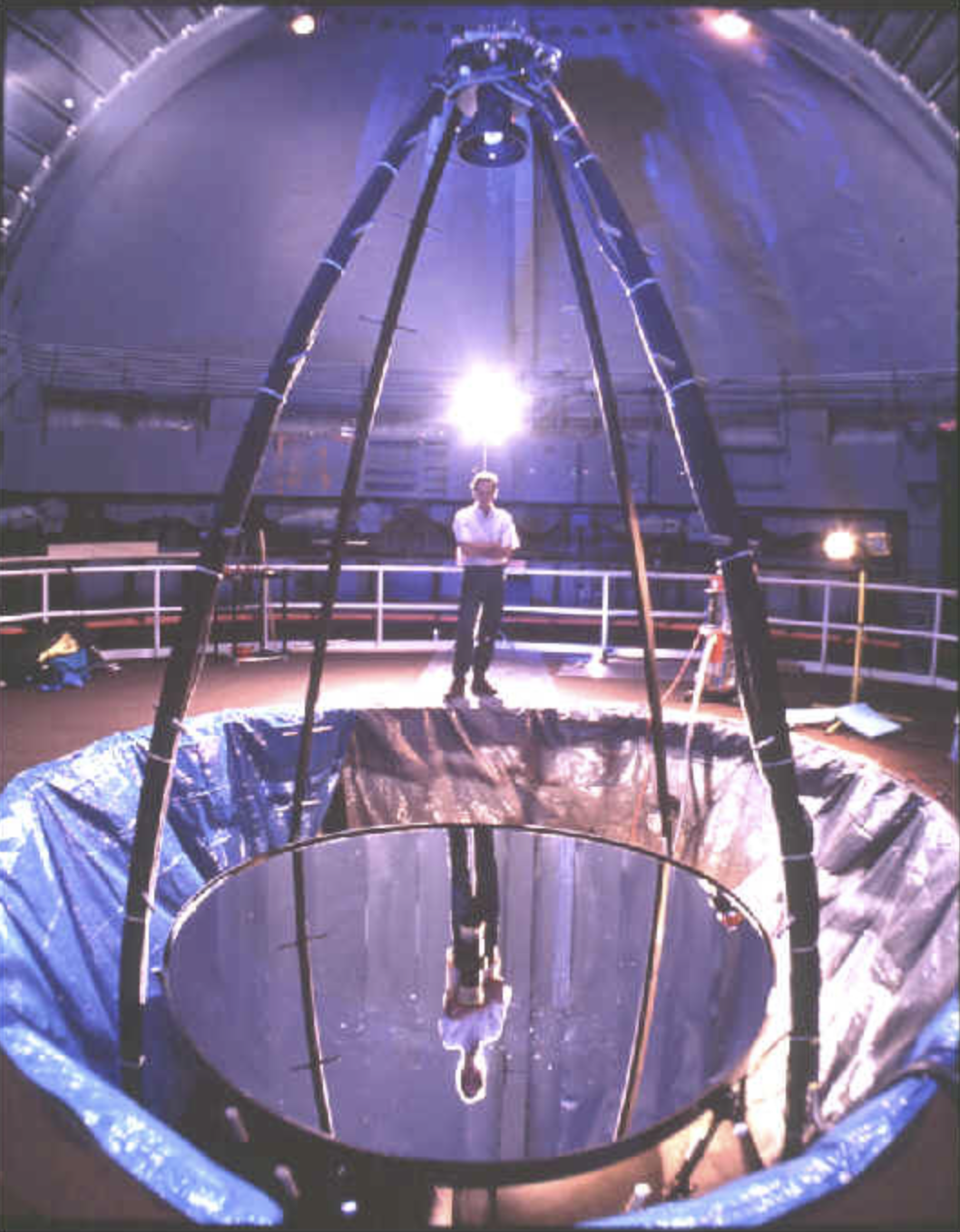}
\caption{The NASA Orbital Debris Observatory (NODO, 1995-2002, \copyright \, Chip Simons Photography). \label{Fig_2}}
\end{minipage}
\hfill
\begin{minipage}{9.0cm}
\includegraphics[width=9.0cm]{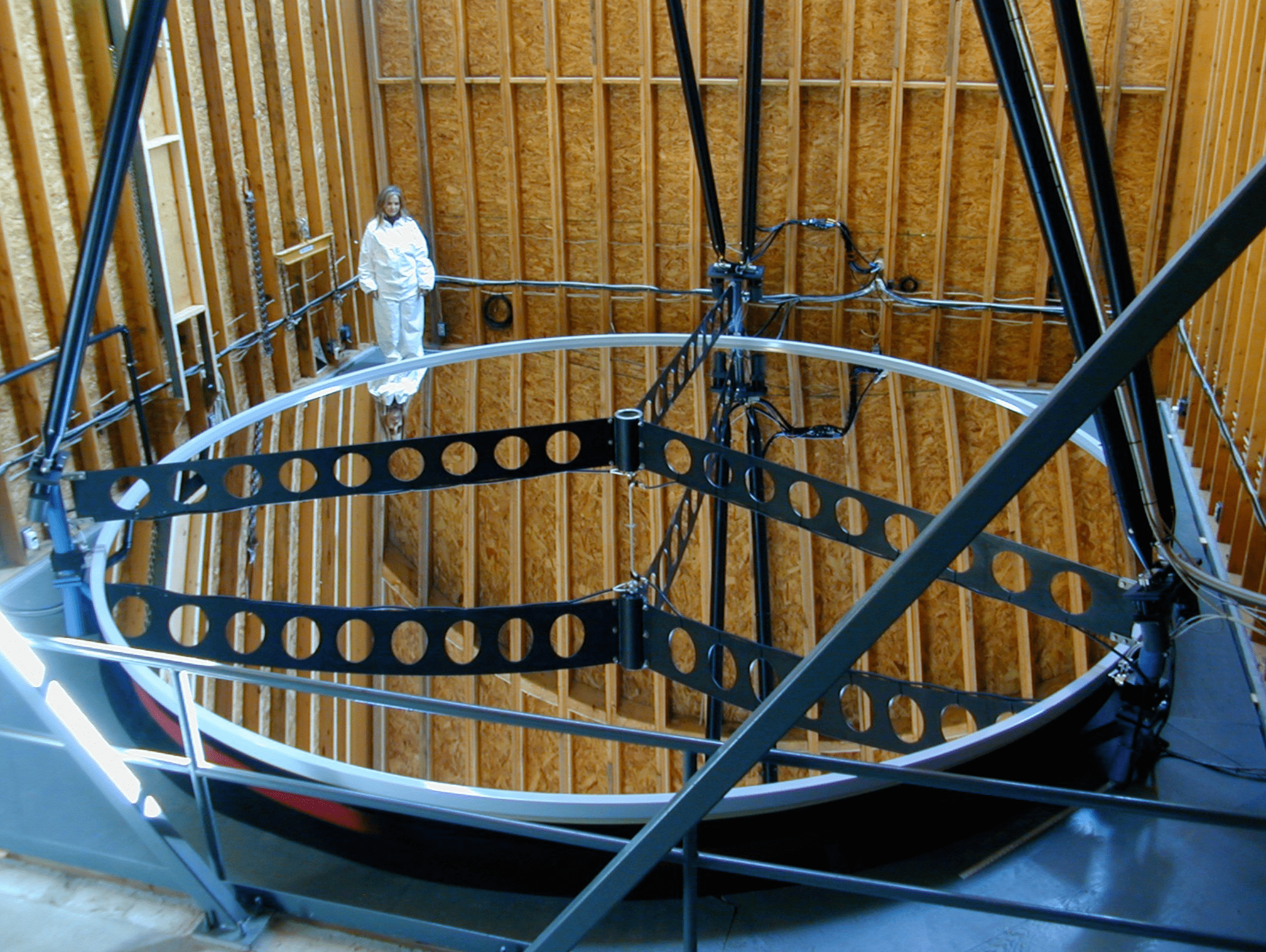}
\caption{The Large Zenith Telescopoe (LZT, 2003-2016, \copyright \, Prof. Paul Hicskon). \label{Fig_3}}
\end{minipage}
\end{figure}

In 1994, Paul Hickson, Ermanno Borra and their colleagues, built an experimental 2.7-m diameter liquid mirror telescope at the University of British Columbia (UBC) near Vancouver (see Fig.~\ref{Fig_1} and \citealt{paper17}). Paul Hickson also worked with NASA on a 3-m liquid mirror telescope in New Mexico to observe space debris (see Fig.~\ref{Fig_2} and \citealt{paper11}).  In the early 2003, Hickson built the 6-m Large Zenith Telescope (LZT) to extend LM technology to larger apertures (see Fig.~\ref{Fig_3} and \citealt{paper2}). Liquid mirrors have also been used by atmospheric scientists for LIDAR applications (\citealt{paper12, paper13}). For instance, the LZT has been used to study with an unprecedented spatial and temporal resolution the structure and dynamics of the Earth's mesosphere and lower thermosphere. The LZT was de-commissioned in 2016.

All these LMTs were first-generation instruments that either were not optimized for astronomy or not located at high-quality astronomical sites. The ILMT has strongly benefited from all the previous technology developments and it has been developed specifically for astronomical research from a good astronomical site.

The ILMT project was first proposed during the {\it Science with LMTs} workshop organized by Prof. Ermanno Borra on April 14-15, 1997 at the Marseille Observatory (France). 

It finally emerged from a scientific collaboration in observational astrophysics between the Li{\`e}ge Institute of Astrophysics and Geophysics (Li{\`e}ge University), several Canadian universities (British Columbia, Laval, Montr{\'e}al, Toronto, Victoria 
and York) and the Aryabatta Research Institute of observational sciencES (ARIES, Nainital, India). Meanwhile, several colleagues from the Royal Observatory of Belgium, the Pozna{\'n} Observatory (Poland), the Ulugh Beg Astronomical Institute of the Uzbek Academy of Sciences and the National University of Uzbekistan, and the Indian Space Research Organization (ISRO) have also joined the ILMT project (\citealt{paper14, paper15}). 

After several years of design work and construction in Belgium by AMOS (Advanced Mechanical and Optical Systems, Li{\`e}ge), CSL (Centre Spatial de Li{\`e}ge), SOCABELEC (Jemeppe-sur-Sambre) and   Li{\`e}ge University, and in India by ARIES, commissioning began in April 2022, and the telescope achieved first light on 29 April 2022 (\citealt{paper16}). The 4-metre ILMT is located on the ARIES site of Devasthal (Uttarakhand, India,  Longitude = $79^{\circ}41'07.08'' E$, Latitude = $29^{\circ}21'41.4'' N$,  Altitude = 2378m). Its inauguration took place on the first Spring day of 2023, i.e. on 21st of March. 

After this brief history on LMs\footnote{A more detailed and general account on the history of LMTs may be found in \citet{paper5}} and the ILMT project, we describe the very special type of optical corrector that is required for such a telescope and we also present some first scientific results.

\section{Need for a special optical corrector}

{\it $^{\prime\prime}$But the fool on the hill,

Sees the Sun going down.

And the eyes in his head,

See the world spinning around \dots $^{\prime\prime}$ }

The Beatles

But if we cannot orient the mirror of our telescope and given that the Earth rotates around its south-north axis, the stars in the focal plane must move $\ldots$ We may naturally wonder what kind of trajectories do the stars follow?

\begin{figure}
\centering
\begin{minipage}{6cm}
\caption{Due to the rotation of the Earth, the light rays coming from the star S sweep a cone which directrix is a circle centred on the south-north axis (dashed black vector) and the generatrix the light ray itself. After entering the optical system at $O$, i.e. the vertex of the cone, the light rays intersect the plane of the CCD detector along a conic, the branch of an hyperbola in the present case, represented by the black bold curve. \label{Fig_4}}
\end{minipage}
\hfill
\begin{minipage}{9cm}
\includegraphics[width=9cm]{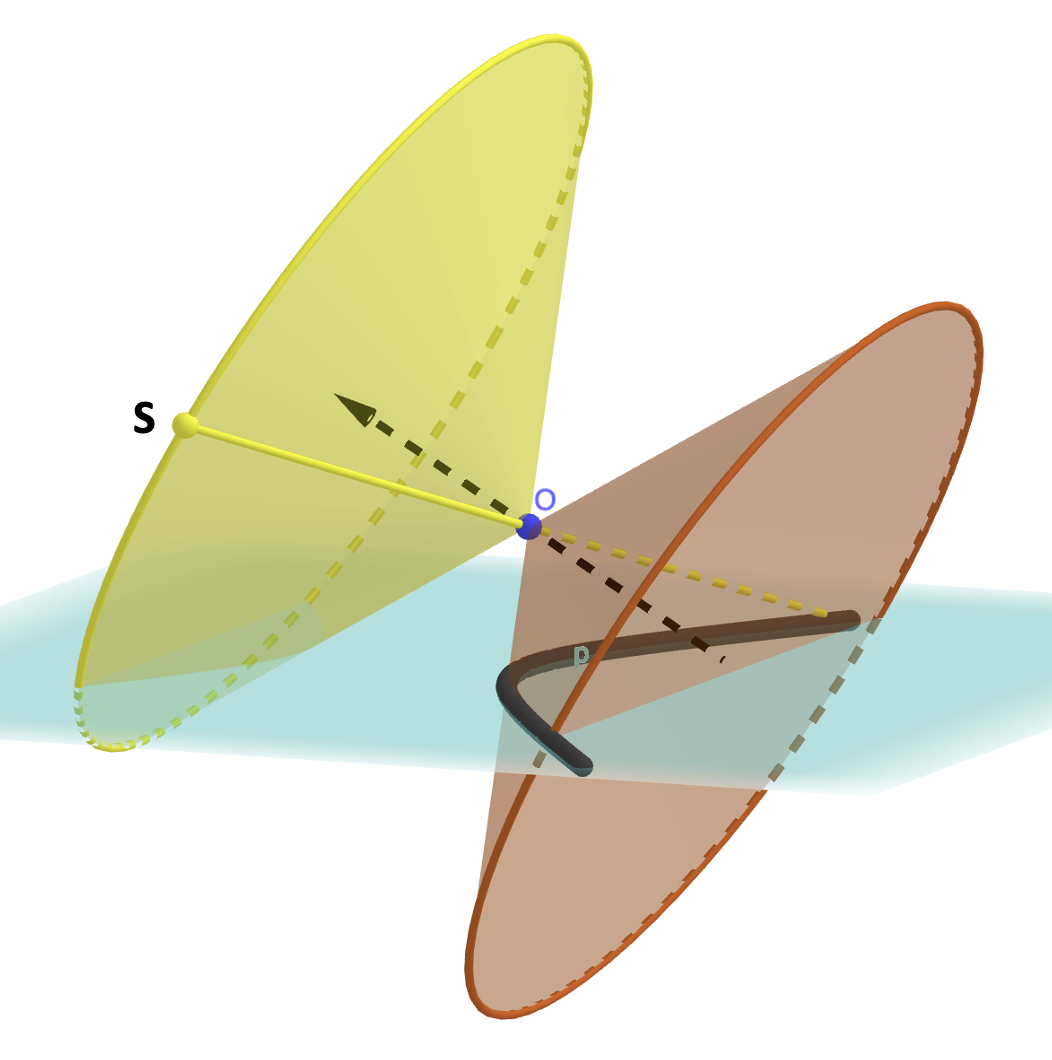}
\end{minipage}
\end{figure}

Since the light rays coming from the stars sweep a conic surface\footnote{degenerated into a plane perpendicular to the south-north axis for the special case of an equatorial star},  the same light rays intersect, after entering an optical system (cf. the aperture of the ILMT), the plane surface of the detector, in the form of a conic trajectory (see Fig.~\ref{Fig_4}).  One may show that for a zenith telescope installed under a latitude of 45$^\circ$, the conic trajectory corresponds to the branch of an hyperbola. It is also the fate of all the solar shadows cast on the ground or on walls to follow conic trajectories during daylight. 

\begin{figure}
\centering
\begin{minipage}{7.5cm}
\includegraphics[width=7.5cm]{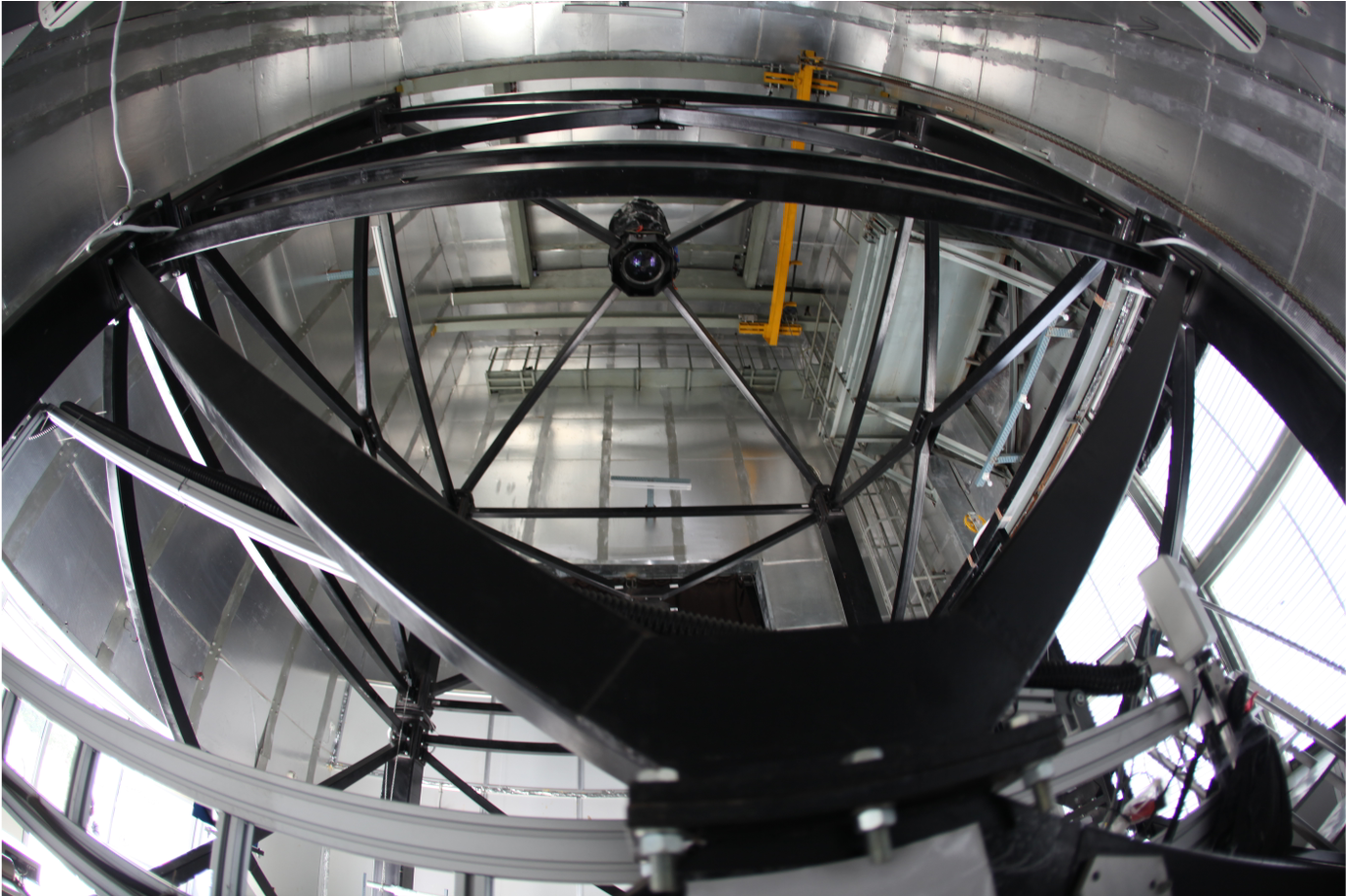}
\caption{Fish eye view from bottom of the main ILMT structure and of the optical corrector at the very top. 
 \label{Fig_5}}
\end{minipage}
\hfill
\begin{minipage}{7.5cm}
\includegraphics[width=7.5cm]{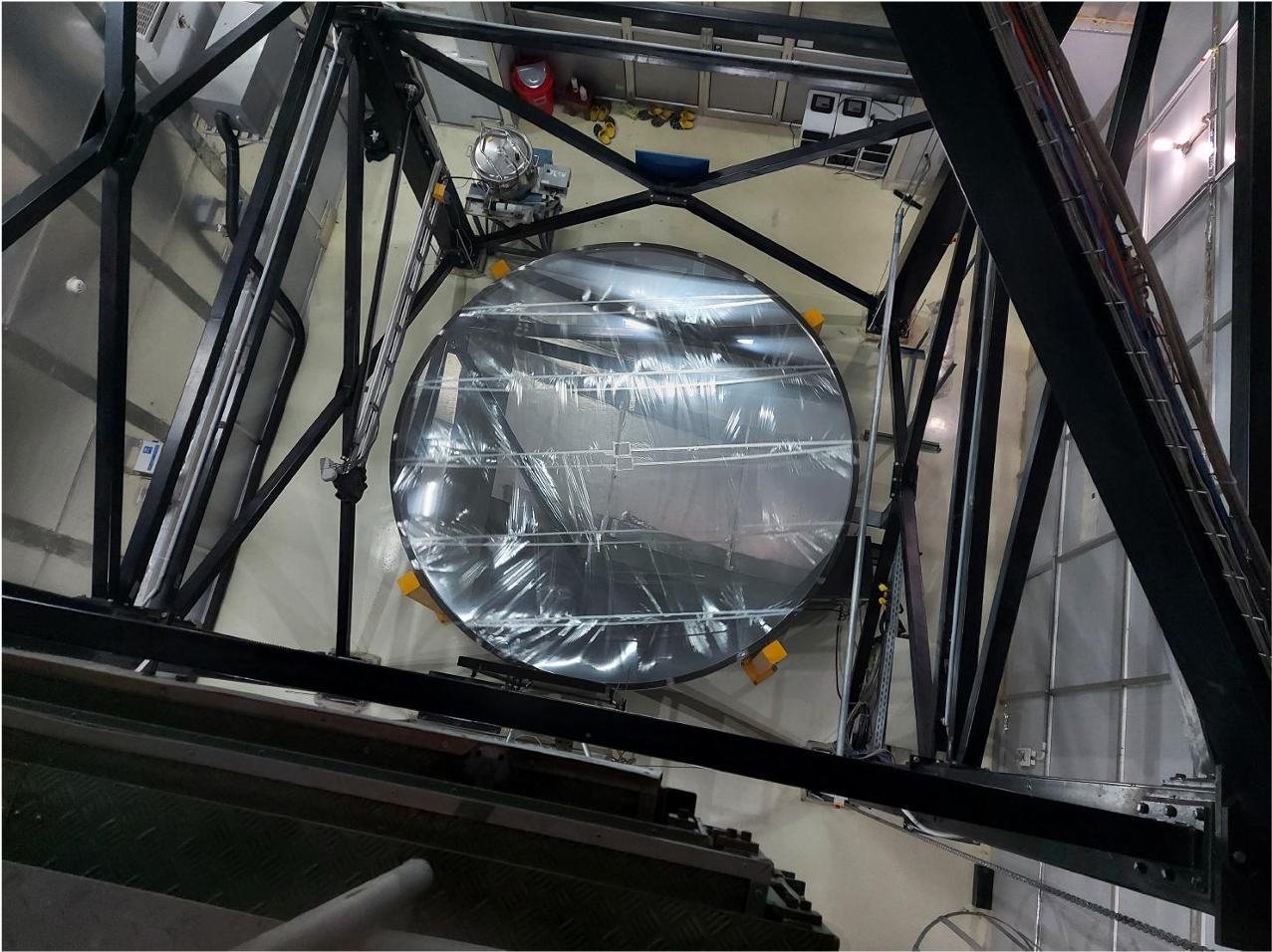}
\caption{Top view of the ILMT mirror filled with mercury and covered with mylar
\label{Fig_6}}
\end{minipage}
\end{figure}

There is thus the need  of a very special optical corrector to be set in front of the CCD to render linear the hyperbolic trajectories of the stars in the focal plane of the ILMT.  Such an optical corrector, named  TDI optical corrector, made of 5 lenses some of which are tilted and/or offset from the optical axis, has been designed by \citet{paper4}.  Fig.~\ref{Fig_5} shows a view of the ILMT taken from bottom. The optical corrector is visible at the top of the structure of the telescope. Fig.~\ref{Fig_6} illustrates the ILMT as seen from the top. One clearly sees the mirror covered with mercury. Several parallel mylar sheets covering the mirror are very conspicuous. They are used to protect the mercury surface from vortices, generated by the rotation of the mirror, in the air above it.   

As previously mentioned, it is possible now to take continuous images of the sky passing overhead, by operating the CCD detector in TDI mode. For the ILMT, the effective exposure time is approximately 102 sec, the time it takes for a star to cross the entire length of the CCD. Due to limitations of the data system, the maximum size of an individual image is $4096\times 40960$ pixels, truncated to $4096\times 36864$ pixels because the first $4096\times 4096$ pixels correspond to the ramping phase of the exposure. Accordingly, there is a small gap between consecutive images while the data are written to disk. The CCD detector is cooled down to a temperature near $-110^{\circ}$C in order to reduce dark current.

\section{First scientific results}

Thanks to the Earth$^\prime$s rotation, the telescope scans a strip of sky centered at a constant declination equal to the latitude of its location  ($+29^{\circ}21^\prime41.4^{\prime\prime}$ for the ARIES Devasthal observatory). As the seasons change, that strip moves in and out of the galactic plane and passes in April very close to the northern galactic pole, an ideal region for the observation of extragalactic objects.  The angular width of the strip is about $22^\prime$, a size limited by that of the CCD detector (4096 $\times$ 4096 pixels) used in the focal plane of the telescope. 

Since the ILMT observes the same region of the sky night after night, it is possible either to co-add the images taken on different nights in order to improve the limiting magnitude or to subtract them from a high S/N reference one to make a variability study of the corresponding strip of sky (\citealt{paper1}). 
Some of us realized in 1997 during the  astronomy workshop in
Marseille that a deep survey of such a narrow strip might lead to the detection of  some 20,000 quasars. Since about one in every one thousand quasars should show multiple images due to gravitational lensing by a foreground galaxy, this could yield of the order of 20 cases of multiply imaged quasars in the ILMT field of view. That would constitute a treasure trove for cosmologists who can use these systems
to study the distribution of dark matter and the geometry and expansion history of the Universe. A cheap and dedicated instrument alike the ILMT would be ideal for such studies (\citealt{paper18}).

\begin{figure}
\centering
\begin{minipage}{8cm}
\caption{A composite image produced from five consecutive nights in October 2022 using successively the Sloan $r$, $r$, $g$, $g$ and $i$ filters. The diagonal streaks were made by the non-active Russian communication satellite, Meridian 3 (37212).  \label{Fig_7}}
\end{minipage}
\hfill
\begin{minipage}{7cm}
\includegraphics[width=7cm]{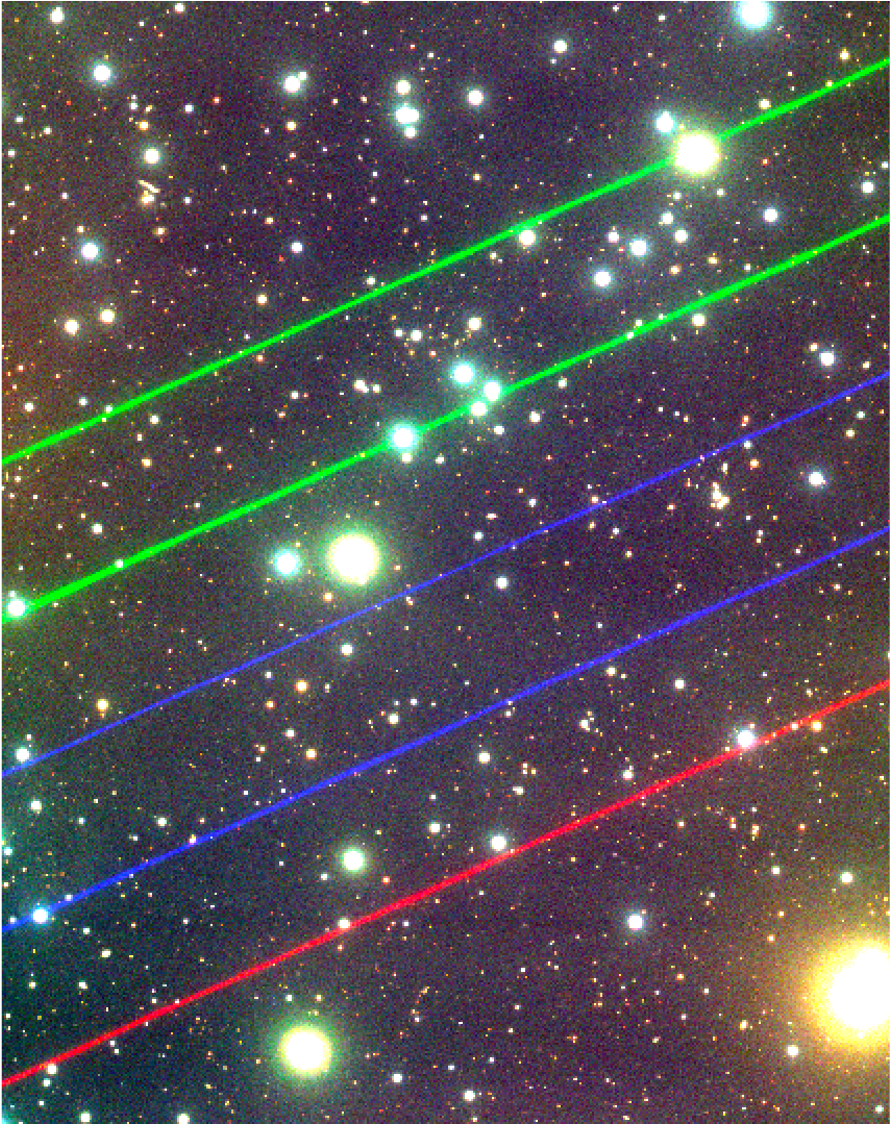}
\end{minipage}
\end{figure}

 Given the large diameter (D = 4m) and comparatively short effective focal length (f = 9.44m) of the ILMT, the latter consists of a powerful survey telescope. Indeed, the detection threshold of any faint moving objects (cf. asteroids, comets, space debris, ...) is proportional to the focal length and inversely proportional to the square of the mirror diameter. Therefore, it is not surprising that more than 80 objects in Earth orbit, among which only about half are cataloged, have been detected during 10 nights of observation with the ILMT during the fall of 2022 (see the ILMT paper by Hickson et al. in these proceedings, and Fig.~\ref{Fig_7}).

\begin{figure}
\centering
\begin{minipage}{7cm}
\caption{Example of an asteroid (3548 Eurybates) observed with the 4m ILMT on 7 consecutive nights in October 2022 using the $g$, $r$ and $i$ Sloan filters. The blue streak is due to a passing satellite.  \label{Fig_8}}
\end{minipage}
\hfill
\begin{minipage}{8cm}
\includegraphics[width=8cm]{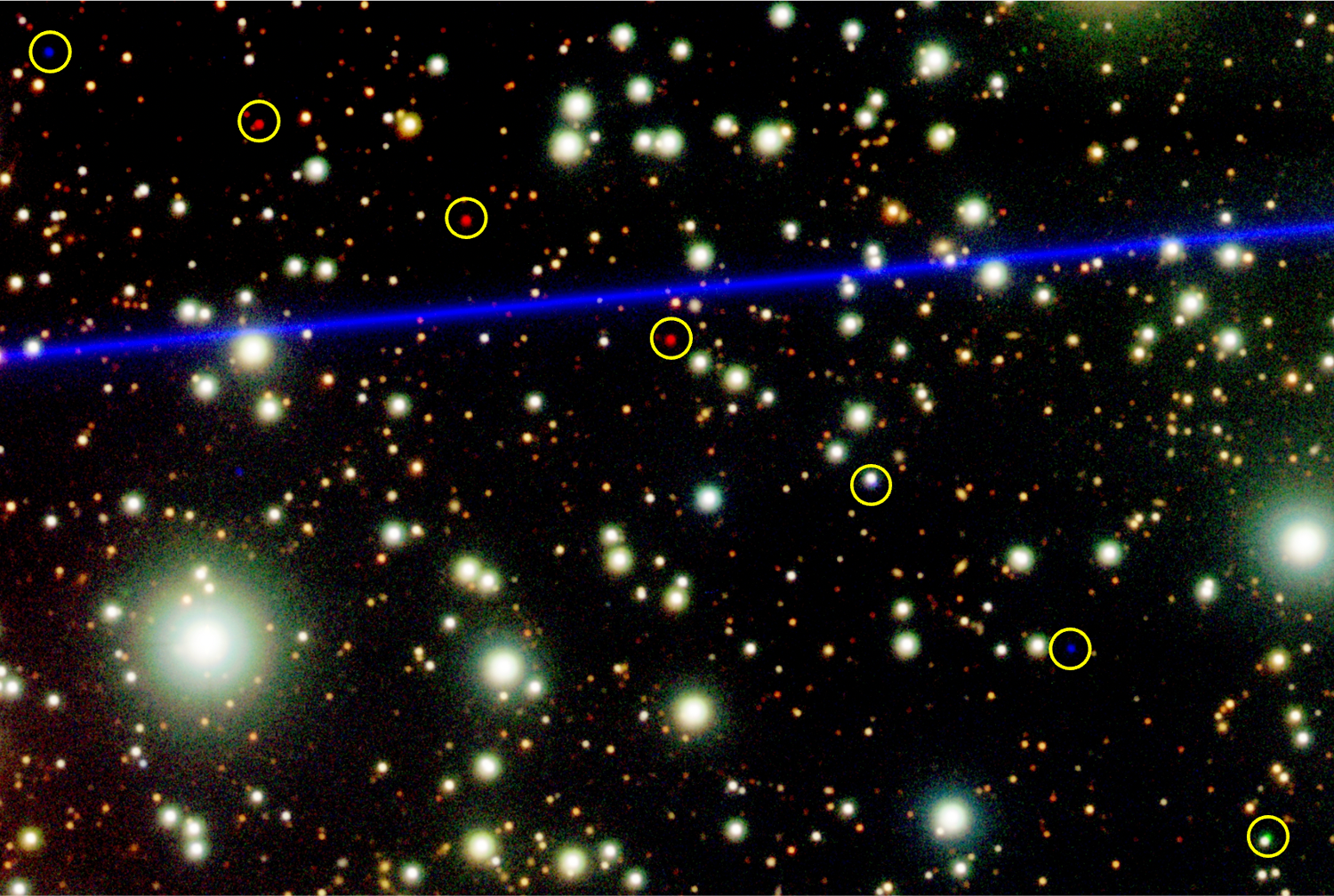}
\end{minipage}
\end{figure}

Similarly, some 75 known asteroids have been detected during 9 consecutive nights in October-November 2022 in just one field which angular extent is $22^{\prime} \times 198^{\prime}$, i.e. covered during $\sim$ 15 minutes (see the ILMT poster paper by Pospieszalska-Surdej et al. in these BINA3 proceedings, and Fig.~\ref{Fig_8}). 

\begin{figure}
\centering
\begin{minipage}{7.15cm}
\includegraphics[width=7.15cm]{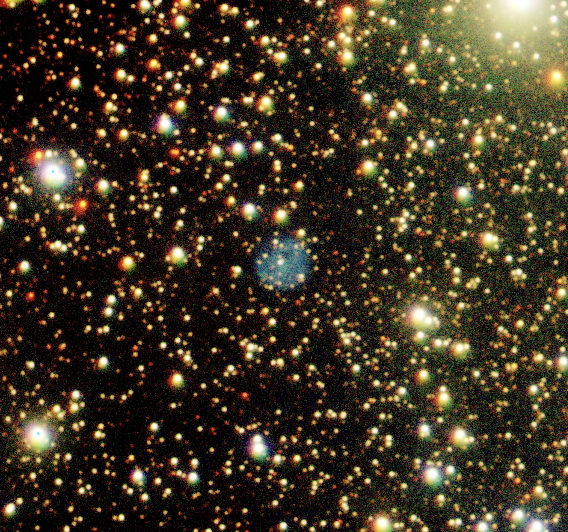}
\caption{Composite $g$, $r$, $i$ image of the planetary nebula NGC 6842 in the field of view of the ILMT. 
 \label{Fig_9}}
\end{minipage}
\hfill
\begin{minipage}{7.9cm}
\includegraphics[width=7.8cm]{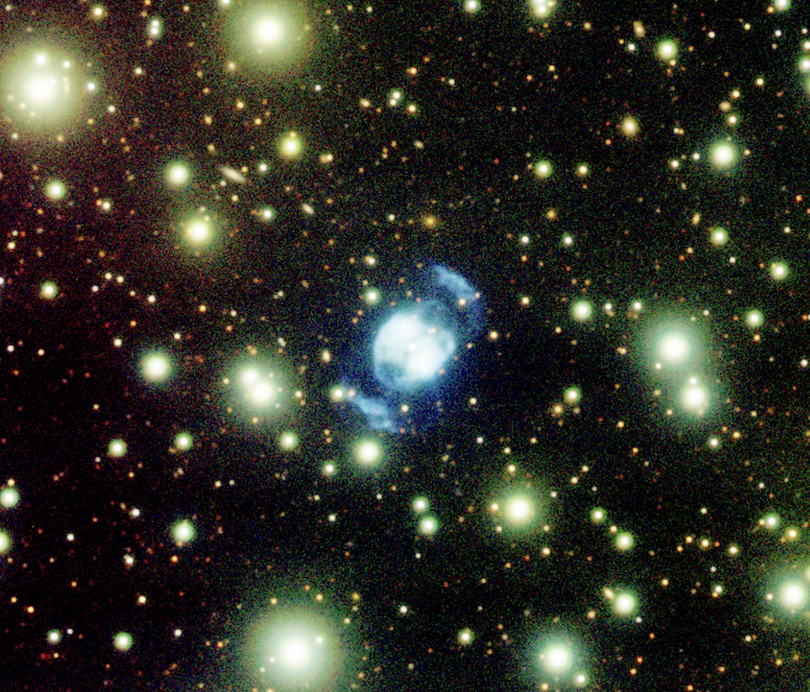}
\caption{Composite $g$, $r$, $i$ image of the planetary nebula NGC 2371 in the field of view of the ILMT. 
\label{Fig_10}}
\end{minipage}
\end{figure}

\begin{figure}
\centering
\begin{minipage}{7.5cm}
\includegraphics[width=7.5cm]{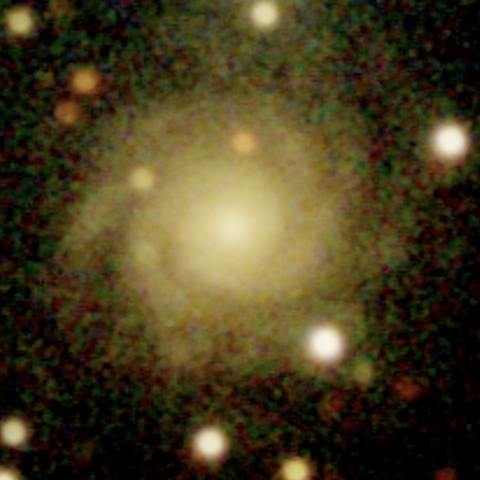}
\caption{Composite $g$, $r$, $i$ image of a field observed with the ILMT containing a nice spiral galaxy.  
 \label{Fig_11}}
\end{minipage}
\hfill
\begin{minipage}{7.5cm}
\includegraphics[width=7.5cm]{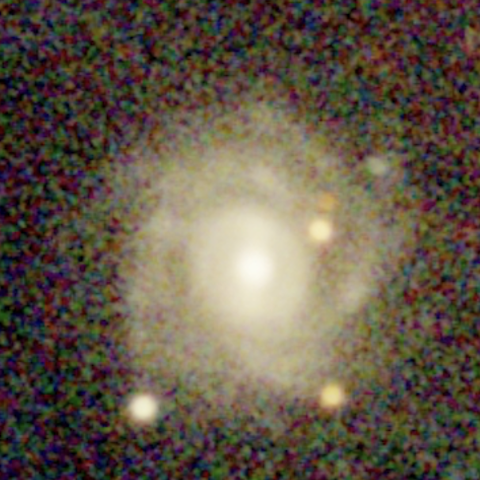}\caption{Composite $g$, $r$, $i$ image of a field observed with the ILMT containing another spiral galaxy.
\label{Fig_12}}
\end{minipage}
\end{figure}

\begin{figure}
\centering
\begin{minipage}{7.5cm}
\includegraphics[width=7.5cm]{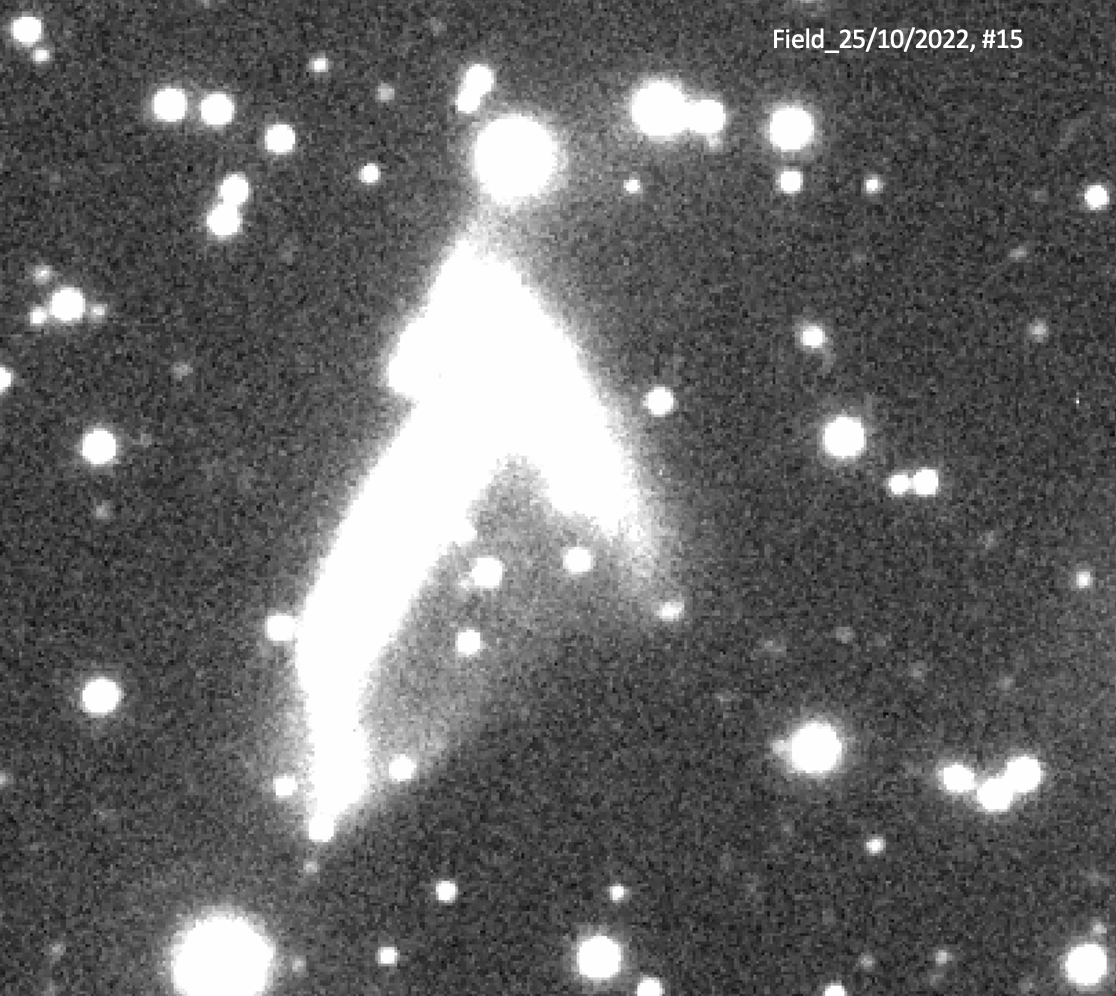}
\caption{ILMT image in the $i$ spectral band of a strange object which physical nature is revealed on the right image after changing its dynamic. 
 \label{Fig_13}}
\end{minipage}
\hfill
\begin{minipage}{7.9cm}
\includegraphics[width=7.8cm]{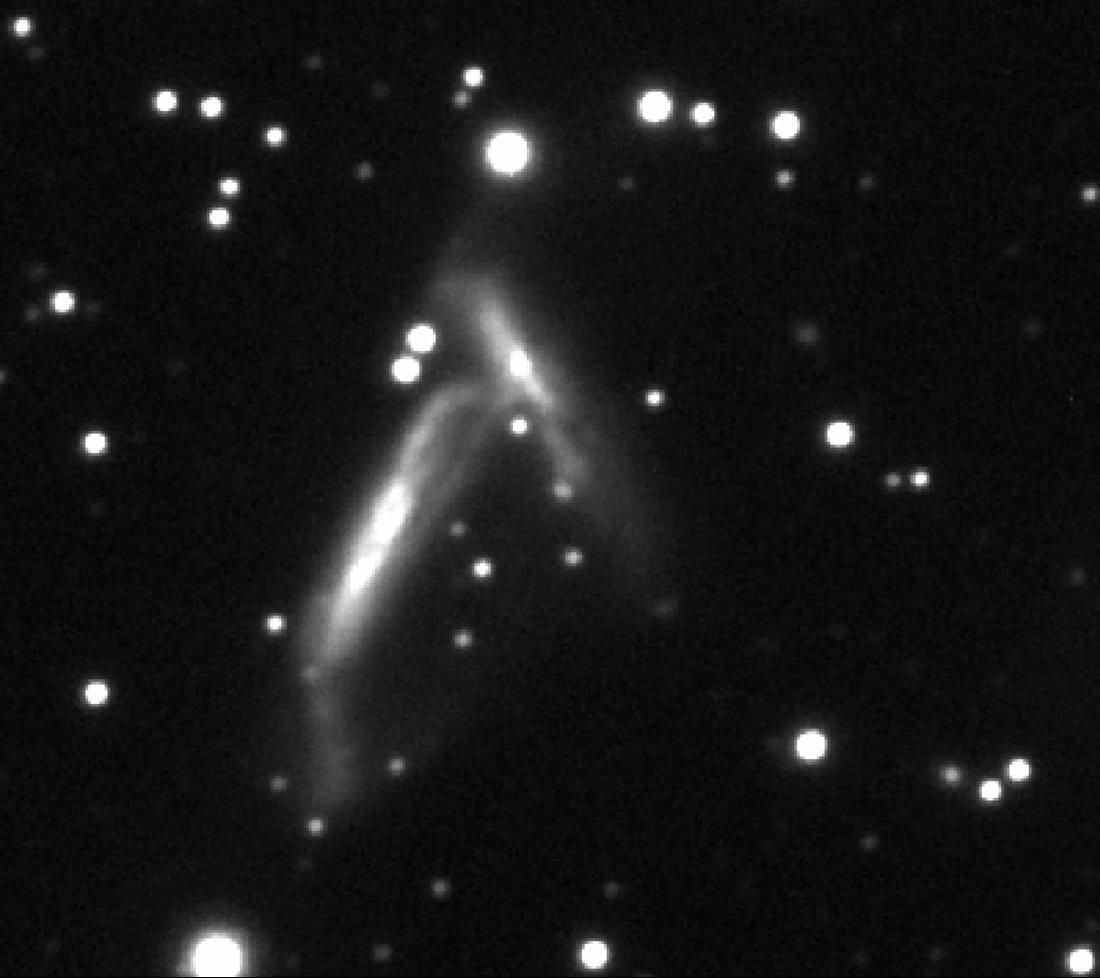}
\caption{Nice system of at least two interacting galaxies with transfer of material between them.
\label{Fig_14}}
\end{minipage}
\end{figure}

\begin{figure}
\centering
\begin{minipage}{7.7cm}
\includegraphics[width=7.7cm]{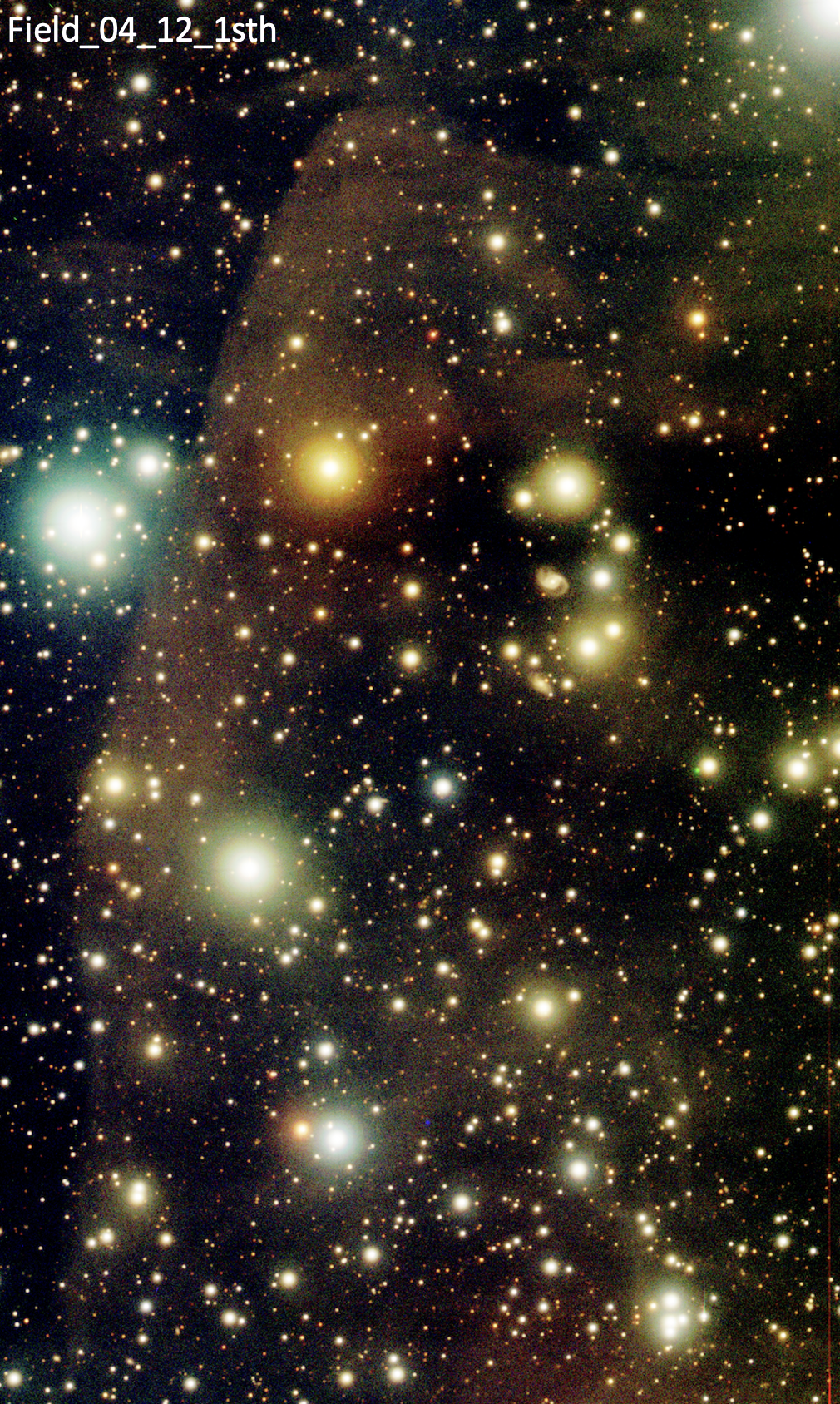}
\caption{ILMT composite $g$, $r$, $i$ image of a reflection nebula in one of the ILMT fields (near LST, i.e. Local Sidereal Time = 04h 12m in October 2022).  
 \label{Fig_15}}
\end{minipage}
\hfill
\begin{minipage}{7.3cm}
\includegraphics[width=7.3cm]{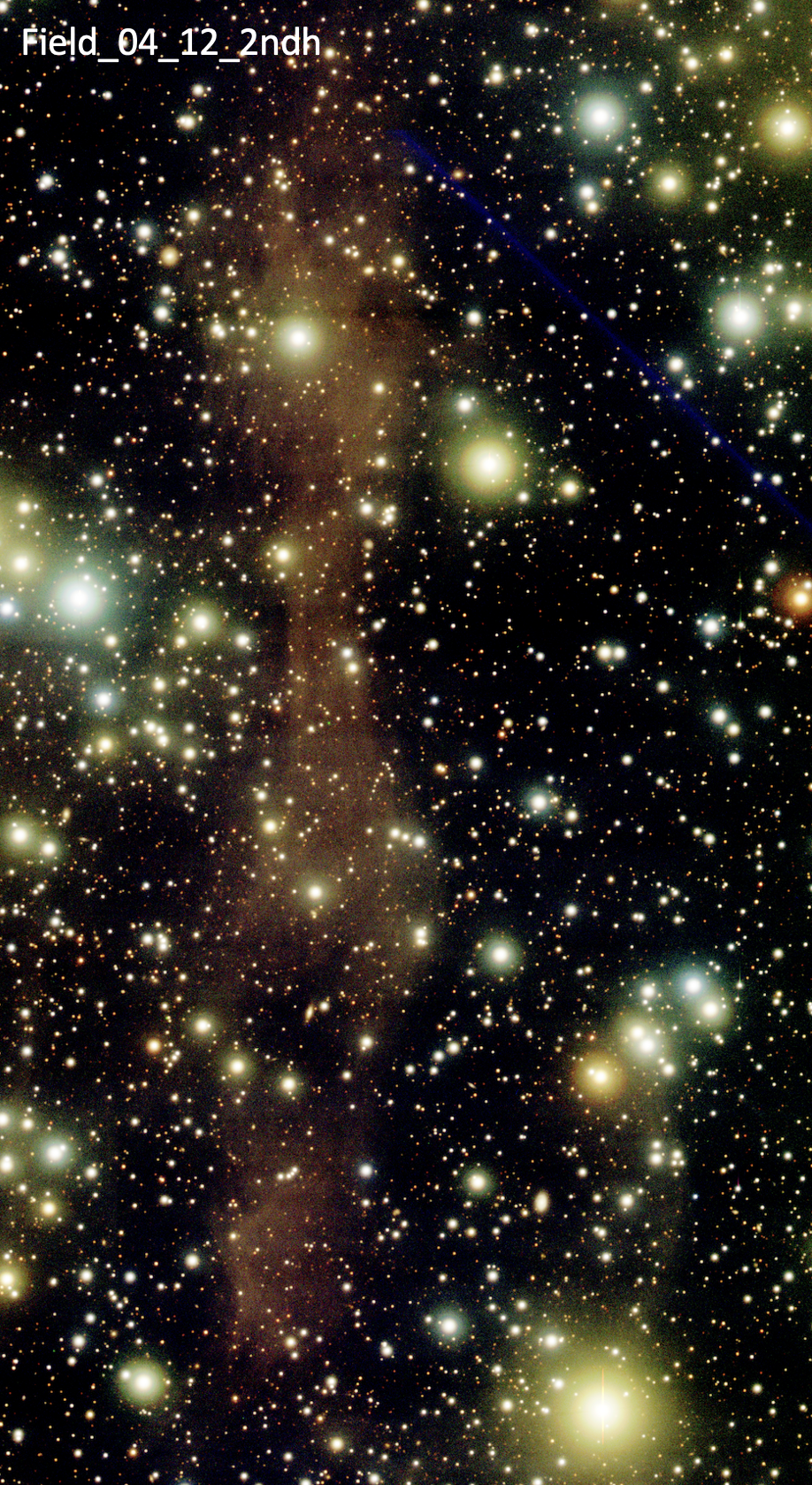}
\caption{ILMT composite $g$, $r$, $i$ image of a reflection nebula in one of the ILMT fields (near LST = 04h12m in October 2022).
\label{Fig_16}}
\end{minipage}
\end{figure}

\begin{figure}
\centering
\begin{minipage}{9.5cm}
\includegraphics[width=9.5cm]{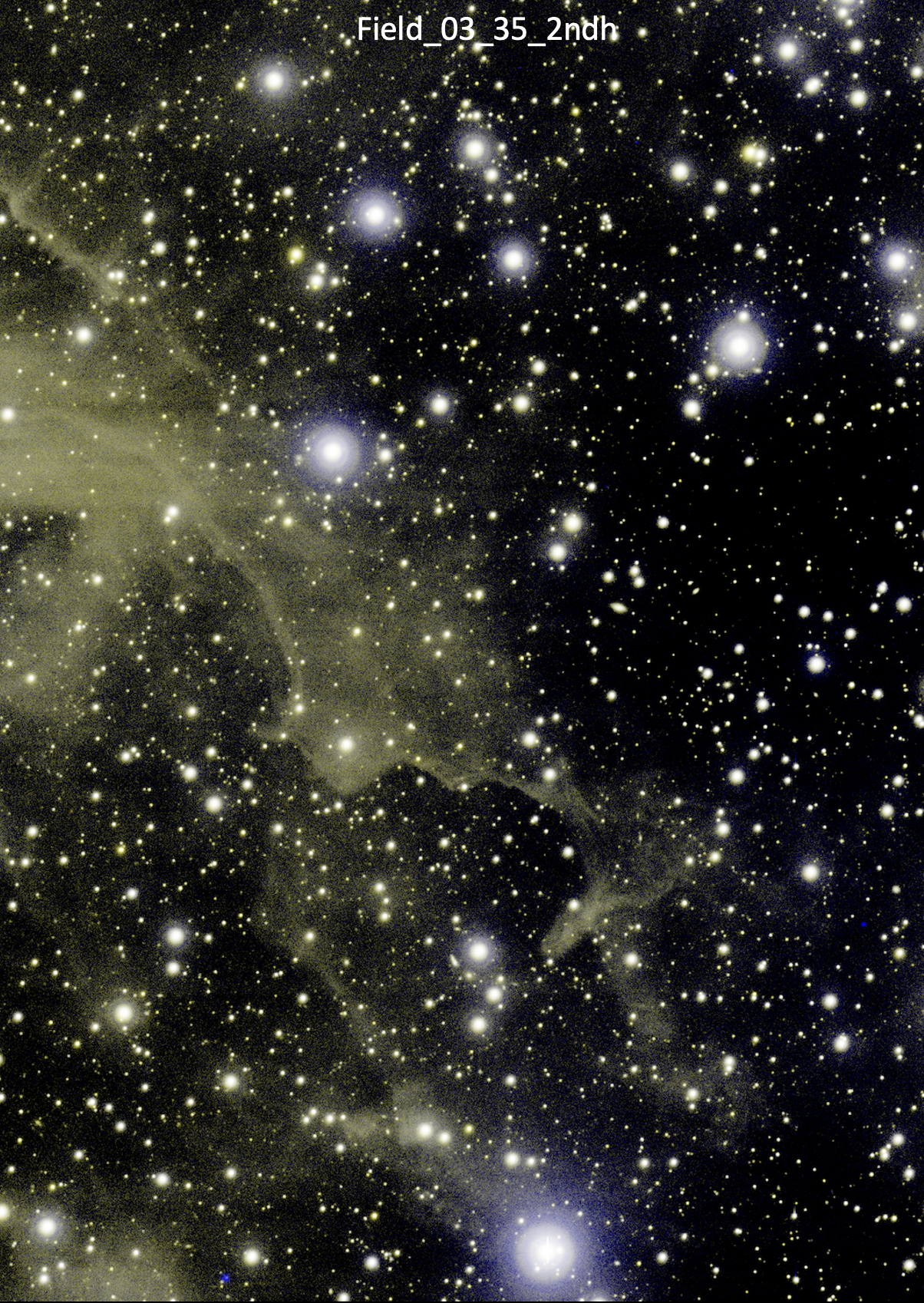}
\caption{ILMT composite $g$, $r$, $i$ image of a reflection nebula in one of the ILMT fields (near LST = 03h 35m in October 2022).  
 \label{Fig_17}}
\end{minipage}
\hfill
\begin{minipage}{5.5cm}
\includegraphics[width=5.5cm]{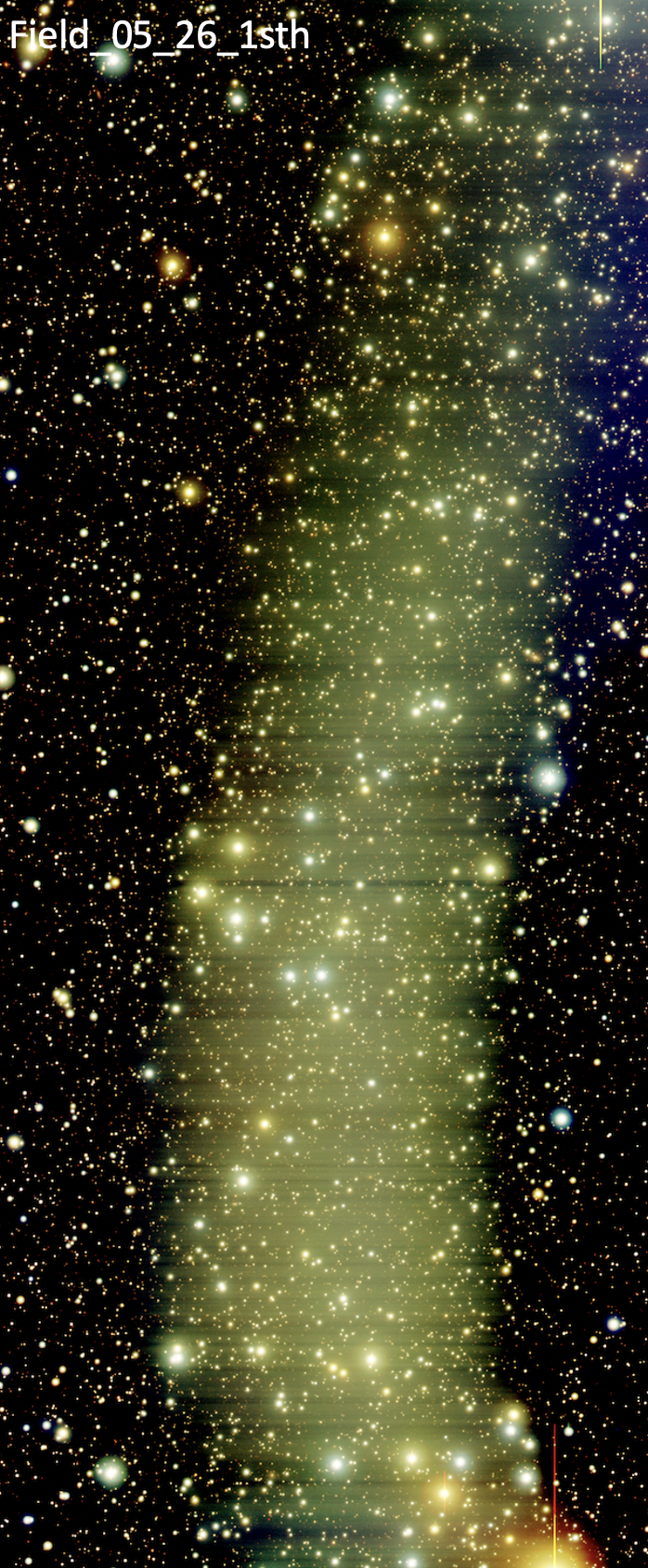}
\caption{ILMT composite $g$, $r$, $i$ image of a reflection nebula in one of the ILMT fields (near LST = 05h26m in October 2022).
\label{Fig_18}}
\end{minipage}
\end{figure}

\begin{figure}
\centering
\includegraphics[width=15cm]{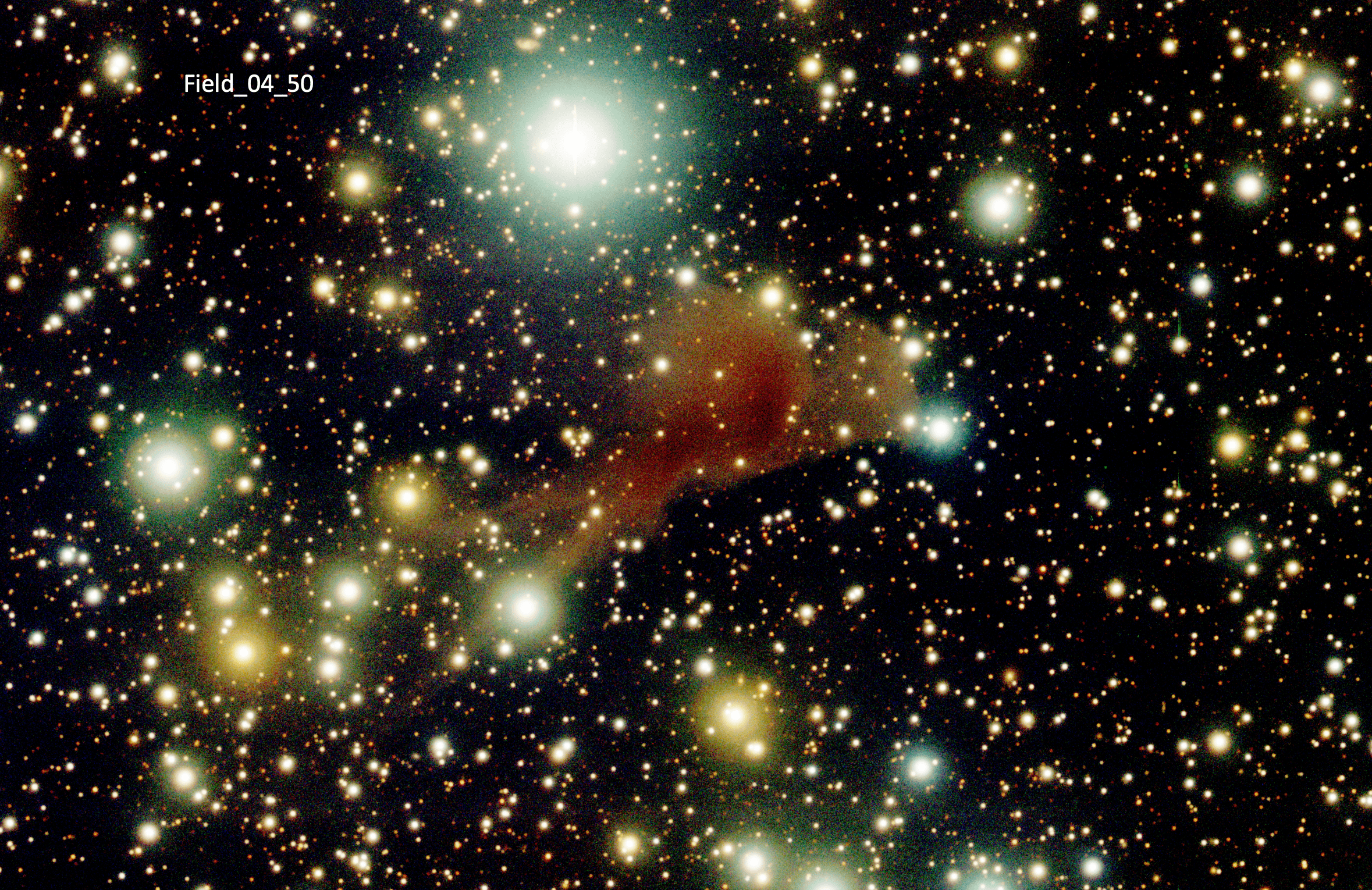}
\begin{minipage}{15.0cm}
\caption{ILMT composite $g$, $r$, $i$ image of a reflection nebula in one of the ILMT fields (near LST = 04h50m in October 2022). \label{Fig_19} }
\end{minipage}
\end{figure}

\begin{figure}
\centering
\begin{minipage}{6.5cm}
\includegraphics[width=6.5cm]{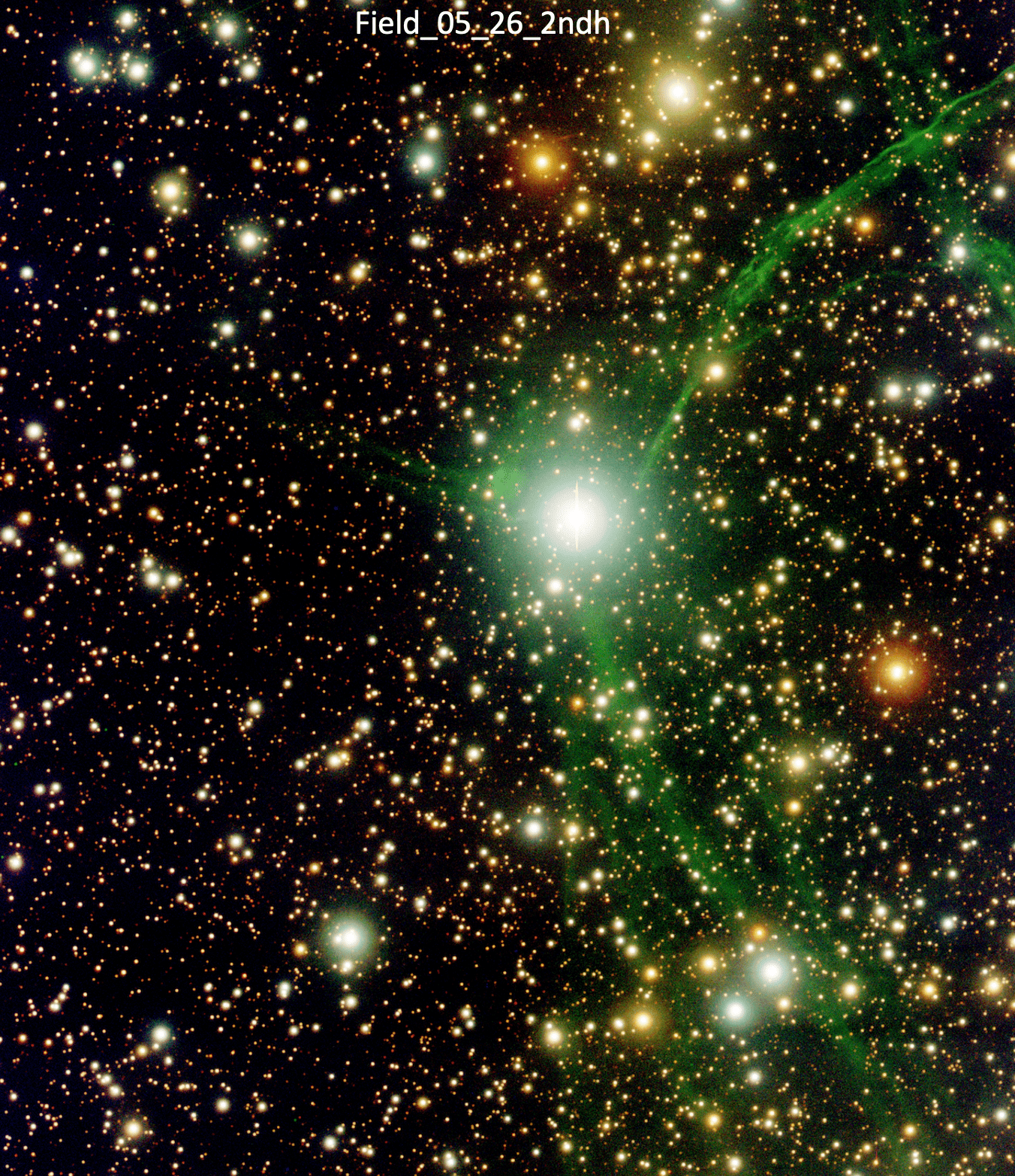}
\caption{ILMT composite $g$, $r$, $i$ image of an emission-line nebula in one of the ILMT fields (near LST = 05h 26m in October 2022).  
 \label{Fig_20}}
\end{minipage}
\hfill
\begin{minipage}{8.5cm}
\includegraphics[width=8.5cm]{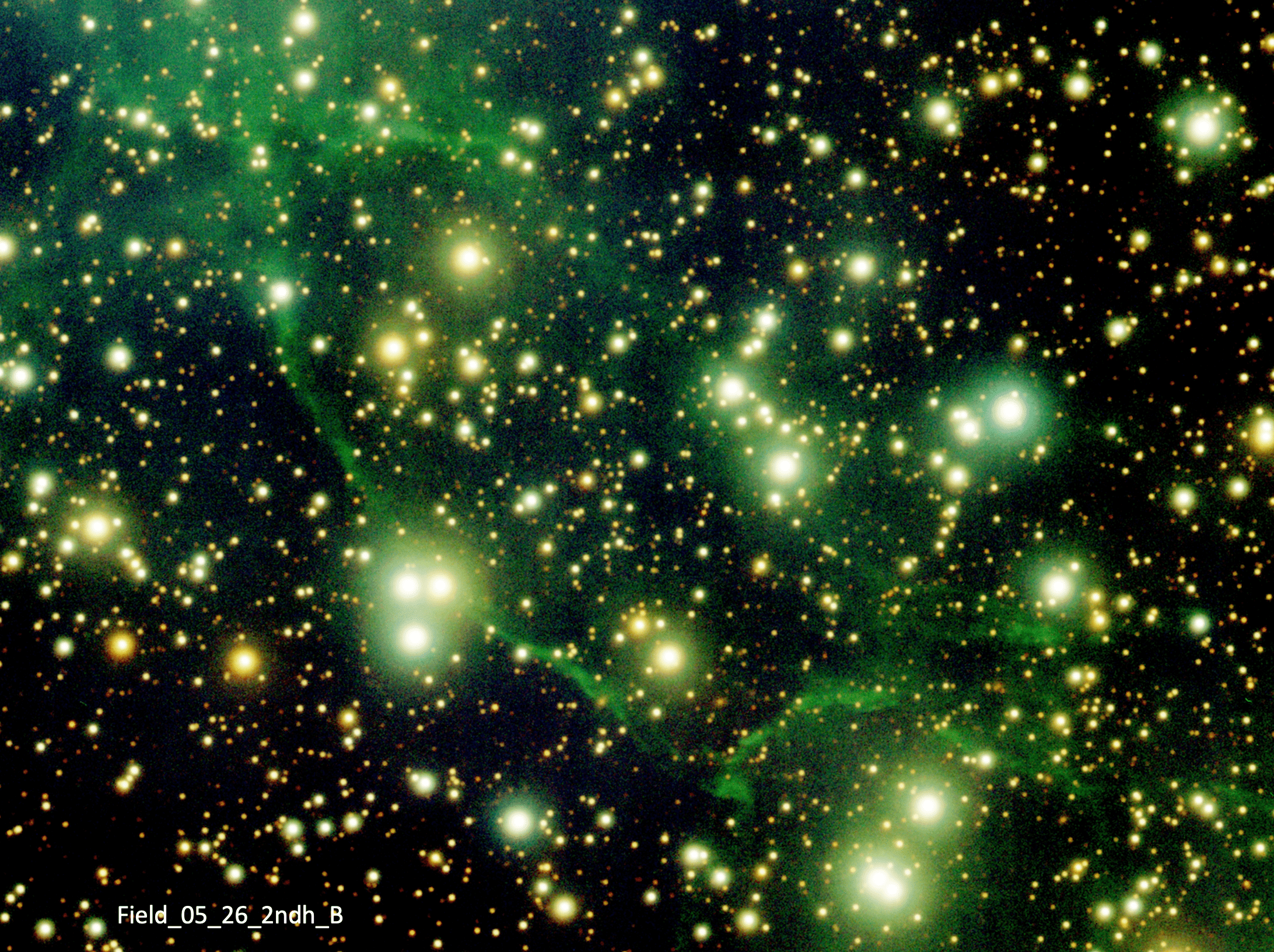}
\caption{ILMT composite $g$, $r$, $i$ image of an emission-line nebula in one of the ILMT fields (near LST = 05h26m in October 2022).
\label{Fig_21}}
\end{minipage}
\end{figure}

Finally, given the fast ratio f/D $\sim$ 2.4 of the ILMT, it constitutes a very sensitive instrument for the detection of low surface brightness objects. Nice examples are given for its capability to detect planetary nebulae (see Figs.~\ref{Fig_9} and~ \ref{Fig_10}), galaxies (see Figs.~\ref{Fig_11} and~\ref{Fig_12}) including the possible appearance of many supernovae (\citealt{paper7, paper8}),  interacting galaxies (see Figs.~\ref{Fig_13} and~\ref{Fig_14}), also some nice reflection nebulae (see Figs.~\ref{Fig_15}-\ref{Fig_19}) as well as emission-line nebulae (see Figs.~\ref{Fig_20}-\ref{Fig_21}). Most of those images are composite frames taken through the broadband ($\Delta \lambda \sim 150$ nm) $g$, $r$ and $i$ Sloan filters centred around the wavelength $\lambda \,\, 468.6$ nm, $616.5$ nm and $748.1$  nm, respecively (\citealt{Fukugita}).


\section{Conclusions}

The ILMT is an instrument that can be entirely dedicated to a photometric and astrometric variability survey, as well as to search for astronomical transients. Over the course of a year, it is able to image 117 square degrees of sky with a pixel size of $0.327^{\prime\prime}$ and an angular resolution that is not limited by the diffraction limit of the telescope but by the natural seeing prevailing at the site (median $FWHM \sim 1.2^{\prime\prime}$).
 
 The main advantages of LMTs are several. There is the possibility of continuous observations at the zenith where the image quality is best, atmospheric extinction the smallest as well as light pollution. It is also remarkable that a 1-D flat field can be constructed from the median signal of the sky background recorded along each column of the CCD. This leads to high quality photometric calibration of the science frames.  

 LMTs are also very much appreciated for their low cost and ease of construction compared to conventional telescopes based upon glass mirrors.
 
 Among the disadvantages is the fact that LMTs are non steerable and that the integration time is limited (102 sec. in the case of the ILMT). But the co-addition of CCD frames acquired with the ILMT during subsequent nights leads to the possible detection of very faint objects. Image subtraction performed on a set of very uniform CCD frames will also reveal the appearance of faint transients. Applications of the ILMT include the detection and photometric follow up of supernovae, multiply imaged quasars, variable stars, asteroids, low-surface-brightness objects, and space debris. 












\begin{acknowledgments}
The 4m International Liquid Mirror Telescope (ILMT) project results from a collaboration between the Institute of Astrophysics and Geophysics (University of Li\`{e}ge, Belgium), the Universities of British Columbia, Laval, Montreal, Toronto, Victoria and York University, and Aryabhatta Research Institute of observational sciencES (ARIES, India). The authors thank Hitesh Kumar, Himanshu Rawat, Khushal Singh and other observing staff for their assistance at the 4m ILMT.  The team acknowledges the contributions of AMOS (Advanced Mechanical and Optical Systems), CSL (Centre Spatial de Li{\`e}ge), SOCABELEC (Jemeppe-sur-Sambre) and ARIES's past and present scientific, engineering and administrative members in the realisation of the ILMT project. JS wishes to thank Service Public Wallonie, F.R.S.-FNRS (Belgium) and the University of Li\`{e}ge, Belgium for funding the construction of the ILMT. PH acknowledges financial support from the Natural Sciences and Engineering Research Council of Canada, RGPIN-2019-04369. PH and JS thank ARIES for hospitality during their visits to Devasthal. B.A. acknowledges the Council of Scientific $\&$ Industrial Research (CSIR) fellowship award (09/948(0005)/2020-EMR-I) for this work. M.D. acknowledges Innovation in Science Pursuit for Inspired Research (INSPIRE) fellowship award (DST/INSPIRE Fellowship/2020/IF200251) for this work. T.A. thanks Ministry of Higher Education, Science and Innovations of Uzbekistan (grant FZ-20200929344).

\end{acknowledgments}

\begin{furtherinformation}

\begin{orcids}
\orcid{0000-0002-7005-1976}{Jean}{Surdej} 
\end{orcids}

\begin{authorcontributions}
This work results from a long-term collaboration to which all authors have made significant contributions.

\end{authorcontributions}

\begin{conflictsofinterest}
The authors declare no conflict of interest.
\end{conflictsofinterest}

\end{furtherinformation}

\bibliographystyle{bullsrsl-en}
\bibliography{extra}
\end{document}